\documentclass[aps,prd,twocolumn,superscriptaddress]{revtex4-1}%
\usepackage[dvipsnames]{xcolor}
\usepackage[utf8]{inputenc}
\usepackage{dsfont}
\usepackage{IEEEtrantools}
\usepackage{amsfonts}
\usepackage{amsmath}
\allowdisplaybreaks[4]        
\usepackage{amssymb}
\usepackage{euscript}     
\usepackage{lipsum}[1]
\usepackage{braket}
\usepackage{starfont}
\usepackage{color,soul}         
\usepackage{tensor}        
\usepackage{amsthm}
\usepackage{graphicx}
\usepackage{slashed}
\usepackage{leftidx}
\usepackage{subfigure}
\usepackage{bbm}
\definecolor{outerspace}{rgb}{0.25, 0.29, 0.3}
\definecolor{scarlet}{rgb}{1.0, 0.13, 0.0}
\usepackage[header,title,page,titletoc]{appendix}  
\definecolor{princetonorange}{rgb}{1.0, 0.56, 0.0}
\definecolor{WildStrawberry}{rgb}{1.0, 0.26, 0.64}
\definecolor{rossocorsa}{rgb}{0.83, 0.0, 0.0}
\definecolor{navyblue}{rgb}{0.0, 0.0, 0.5}
\usepackage{float}
\usepackage{mathtools}

\DeclareMathAlphabet{\pazocal}{OMS}{zplm}{m}{n}
\newcommand{\req}[1]{(\ref{#1})} %{Eq.\thinspace(\ref{#1})}  

\newcommand{\bea}{\begin{eqnarray}}

\newcommand{\eea}{\end{eqnarray}}
\newcommand{\ba}{\begin{eqnarray}}
\newcommand{\ea}{\end{eqnarray}}

\newcommand{\be}{\begin{equation}}
\newcommand{\ee}{\end{equation} }
\newcommand{\beqa}{\begin{eqnarray}}
\newcommand{\eeqa}{\end{eqnarray}}
\newcommand{\beqar}{\begin{eqnarray*}}
\newcommand{\eeqar}{\end{eqnarray*}}

\renewcommand{\req}[1]{eq.~(\ref{#1})}

\newcommand{\ssc}{\scriptscriptstyle}
\newcommand{\eg}{{\it e.g.,}\ }
\newcommand{\ie}{{\it i.e.,}\ }

\newcommand{\ctt}{C_{\ssc T}}

\newcommand{\cTFb}[1]{\left.\frac{C_{\ssc T}}{F_0}\right|_{#1}}
\newcommand{\cTb}[1]{\left.C_{\ssc T}\right|_{#1}}
\newcommand{\Fb}[1]{\left.F_0\right|_{#1}}

\newcommand{\iu}{\text{i}}
\newcommand{\expp}[1]{\text{e}^{#1}}

\makeatletter
\newcommand{\dal}{\mathop{\mathpalette\dal@\relax}}
\newcommand{\dal@}[2]{%
  \begingroup
  \sbox\z@{$\m@th#1\square$}%
  \dimen0=\fontdimen8
    \ifx#1\displaystyle\textfont\else
    \ifx#1\textstyle\textfont\else
    \ifx#1\scriptstyle\scriptfont\else
    \scriptscriptfont\fi\fi\fi3
  \makebox[\wd\z@]{%
    \hbox to \ht\z@{%
      \vrule width \dimen0
      \kern-\dimen0
      \vbox to \ht\z@{
        \hrule height \dimen0 width \ht\z@
        \vss
        \hrule height 2\dimen0
      }%
      \kern-2.5\dimen0
      \vrule width 2.5\dimen0
    }%
  }%
  \endgroup
}
\makeatother

\usepackage{hyperref}
\hypersetup{
    colorlinks,
    citecolor=Cyan,
    filecolor=Cyan,
    linkcolor=rossocorsa,
    urlcolor=Cyan
}
\begin{document}

\title{Conformal bounds in three dimensions from entanglement entropy}
\author{Pablo Bueno}
\email{pablobueno@ub.edu}
\affiliation{Departament de F\'isica Qu\`antica i Astrof\'isica, Institut de Ci\`encies del Cosmos Universitat de Barcelona, Mart\'i i Franqu\`es 1, E-08028 Barcelona, Spain}

\author{Horacio Casini}
\email{casini@cab.cnea.gov.ar}
\affiliation{Instituto Balseiro, Centro At\'omico Bariloche, 8400-S.C. de Bariloche, R\'io Negro, Argentina}
%\affiliation{Perimeter Institute for Theoretical Physics, Waterloo, ON N2L 2Y5, Canada}

\author{Oscar Lasso Andino}
\email{oscar.lasso@udla.edu.ec}
\affiliation{Escuela de Ciencias F\'isicas y Matem\'aticas,Universidad de Las Am\'ericas, Redondel del ciclista, Antigua v\'ia a Nay\'on, C.P. 170504, Quito, Ecuador}

\author{Javier Moreno} 
\email{jmoreno@campus.haifa.ac.il}
\affiliation{Department of Physics and Haifa Research Center for Theoretical Physics and Astrophysics, University of Haifa, Haifa 31905, Israel}

\affiliation{Department of Physics, Technion, Israel Institute of Technology,
Haifa, 32000, Israel}

%\date{\today}

\begin{abstract}

The entanglement entropy of an arbitrary spacetime region $A$ in a three-dimensional conformal field theory (CFT) contains a constant universal coefficient, $F(A)$. For general theories, the value of $F(A)$ is minimized when $A$ is a round disk, $F_0$, and in that case it coincides with the Euclidean free energy on the sphere. We conjecture that, for general CFTs, the quantity $F(A)/F_0$ is bounded above by the free scalar field result and below by the Maxwell field one. We provide strong evidence in favor of this claim and argue that an analogous conjecture in the four-dimensional case is equivalent to the Hofman-Maldacena bounds. In three dimensions, our conjecture gives rise to similar bounds on the quotients of various constants characterizing the CFT. In particular, it implies that the quotient of the stress-tensor two-point function coefficient and the sphere free energy satisfies $C_{ \ssc T} / F_0 \leq 3/ (4\pi^ 2 \log 2 - 6\zeta[3]) \simeq 0.14887$ for general CFTs. We verify the validity of this bound for free scalars and fermions, general $O(N)$ and Gross-Neveu models, holographic theories, $\mathcal{N}=2$ Wess-Zumino models and  general ABJM theories.

%Some of the new regular black holes are achieved without requiring any kind of fine tuning between the action parameters and the physical charges    %\comment{...}

%Possible sections: 1. intro, 2. we present formula, 3. we perform general checks for $F(0)\propto a^*$ and for $F''(0)\propto \ctt$ in $d=3,5$ ($D=4,6$) and extend the result to general d.  4. examples, we apply it to various cases (known and unknown). 5. possible generalization of Casini Huerta Myers relation between EE universal coefficient across round spheres $\mathbb{S}^{d-2}$ and free energy of theory on $\mathbb{S}^d$

\end{abstract}
\maketitle

%Higher  curvature  interactions  appear  generically  in string theoretic models, e.g., as $α'$ corrections in the low- energy  effective  action  [26].   However,  rather  than  constructing explicit top-down holographic models, our approach will be to examine simple toy holographic models involving higher curvature gravity in the bulk.  Our perspective is that if there are interesting universal properties which hold for all CFTs, then they should also appear in the holographic CFTs defined by these toy models as well.  This approach has been successfully applied before, e.g., in the discovery of the F-theorem [2, 3].

%{\bf Introduction.} 

The ratio of the trace-anomaly coefficients characterizing any unitary conformal field theory (CFT) in four dimensions is bounded, above and below, by the free scalar and Maxwell field results, respectively \cite{Hofman:2008ar,Hofman:2016awc}
%Unitary conformal field theories (CFTs) in four dimensions satisfy the so-called conformal-collider or Hofmam-Maldacena bounds 
\begin{equation}\label{HMbounds}
 \left. \frac{c}{a}\right|_{\rm Maxwell}\leq  \frac{c}{a} \leq \left. \frac{c}{a}\right|_{\rm free\, scalar}  \, , %=\frac{18}{31}\, ,
\end{equation}
the numerical values being  $18/31$ and $3$ respectively. Roughly, these ``Hofman-Maldacena" (HM) bounds follow from imposing the positivity of the energy flux escaping at null infinity for  states resulting from a local insertion of the stress tensor on the vacuum. Analogous constructions in general spacetime dimensions $d\geq 3$ give rise to constraints involving correlators of the stress tensor \cite{Buchel:2009sk,Chowdhury:2012km}. %In $d=4$, the relevant coefficients turn out to be related to $a$ and $c$ in a way such that \req{HMbounds} follows.

For odd-dimensional CFTs there is no trace anomaly and the coefficients $a,c$ are not defined. A somewhat canonical general-dimension version of $c$ is provided by the stress-tensor two-point function coefficient, $\ctt$, which is proportional to $c$ in $d=4$. On the other hand, a generalization of $a$ which departs from stress-tensor correlators follows from  
% There exist, however, somewhat natural versions of them. They correspond, respectively, to the stress-tensor two-point function coefficient, $\ctt$, and to
 the entanglement entropy (EE) universal coefficient across a round (hyper-)spherical entangling surface, which we denote by $F_0$. Again, in $d=4$ one finds $F_0\propto a$, and hence the analogy. In odd-dimensional theories, this quantity coincides with the Euclidean free energy on the round sphere, $F_0= - \log Z_{\mathbb{S}^d}$ \cite{Dowker:2010yj,CHM}. Also,  in $d=3$ it defines a renormalization group monotone for general theories \cite{Casini:2012ei,Myers:2010xs,Klebanov:2011gs}. 
  
In this letter we present strong evidence that the quotient $\ctt/F_0$ satisfies bounds analogous to (\ref{HMbounds}) for general three-dimensional CFTs, namely,
 \begin{equation}\label{ff04}
  0\leq \frac{C_{\ssc T}}{F_0} \leq \left. \frac{C_{\ssc T}}{F_0}\right|_{\rm free\, scalar}= \frac{3}{ 4\pi^2\log 2- 6\zeta[3]} \simeq  0.14887\dots % \left. \frac{F(A)}{F_0}\right|_{\rm Maxwell} 
   \end{equation}
These are particular cases of more general conjectural bounds involving the EE of arbitrary regions in $d=3$.  %arises from more general considerations involving EE
Given some entangling region $A$, this is given, for a general CFT, by $S^{3d}(A)=c_0 \cdot { \rm perimeter}(\partial A)/\delta- F(A)+\mathcal{O}(\delta)$,
%They appear in EE universal terms. Recently, $F$ minimized by $F_0$ for general CFTs in $d=3$. Makes sense to normalize by $F_0$. \\
%\comment{EE in 3d; $F_0 $ sphere free energy}
%\begin{equation}
%S^{3d}(A)=c_0  \frac{{ \rm perimeter}(\partial A)}{\delta}- F(A)+\mathcal{O}(\delta)\, ,
%\end{equation}
where $c_0$ is a non-universal coefficient, $\delta$ is a UV cutoff, and $F(A)$ is a dimensionless universal coefficient of non-local nature. Naturally, the round-disk case anticipated above corresponds to  $F_0\equiv \left. F \right|_{\partial A=\mathbb{S}^1} $. %This coincides, for general CFTs, with the Euclidean free energy on the round sphere, \ie $F_0= - \log Z_{\mathbb{S}^3}$ \cite{Dowker:2010yj,CHM}. Also, it defines a renormalization group monotone for general theories \cite{Casini:2012ei}. Additionally,
Recently, it has been proved in  \cite{Bueno:2021fxb} that $F_0$ minimizes $F(A)$ for any given CFT, namely,
$
F(A)/F_0 \geq 1$ with $F(A)=F_0 \, \Leftrightarrow A=$ round disk. Consequently, $F_0$ provides a canonical normalization for $F(A)$.  
%The conjecture that represents the central proposal of this letter reads as follows.
With these provisos in mind, we are ready to formulate the conjecture which is the central proposal of this letter.
%Without further ado, let us present the conjecture which is the central proposal of this letter. 
 \begin{itemize}
 \item[{\footnotesize $\blacklozenge$}] {\bf Conjecture:} for general CFTs in three dimensions, the universal coefficient in the entanglement entropy, $F(A)$, of an arbitrary region $A$ normalized by the disk result, $F_0$, is bounded above by the free scalar result and below by the free Maxwell field one. Namely, we conjecture that
 \begin{equation}\label{ff0}
    \left. \frac{F(A)}{F_0}\right|_{\rm Maxwell} \leq  \frac{F(A)}{F_0} \leq   \left. \frac{F(A)}{F_0}\right|_{\rm free\, scalar} 
   \end{equation}
holds for general entangling regions and CFTs.
 \end{itemize}
The lower bound is in fact equivalent to the number $n$ of connected components in the boundary of $A$, and was proved in \cite{Bueno:2021fxb}. The Maxwell field saturates the lower bound and so do topological theories.
 
The rest of the letter will be devoted to provide evidence in support of the above conjecture and to extract various consequences.

% {\bf Universal bounds on the $F$ term: } 
%In this first part we gather evidence in favour of the conjecture (\ref{ff0}).

{\bf Hints from four dimensions:} In $d=4$, the EE universal term is local in nature and appears as the coefficient of a logarithmic divergence. It  is given by a linear combination of two theory-independent local integrals over the corresponding entangling surfaces, which appear weighted by the trace-anomaly coefficients $a,c$. The expression reads \cite{Solodukhin:2008dh,Perlmutter:2015vma}
\begin{equation}
\frac{S^{4d}_{\rm log}(A)}{a} = \frac{1}{\pi } \left[  \mathcal{W}_{\partial A}  + \left( \frac{c}{a}-1\right) \frac{\mathcal{K}_{\partial A} }{2}\right] \, ,
\end{equation}
where $ \mathcal{W}_{\partial A} $ is the so-called Willmore energy \cite{willmore1996riemannian} of $\partial A$  and $\mathcal{K}_{\partial A}$ is an integral involving a quadratic combination of extrinsic curvatures of $\partial A$. Observe that we normalized the expression by $a$ following the analogy with the three-dimensional case \footnote{As opposed to the round disk in $d=3$, the sphere is the  EE universal term extremizer for general $d=4$ theories only within the class of regions with genus $g=0,1$ \cite{Astaneh:2014uba}, but not for $g\geq 2$. \cite{Perlmutter:2015vma}. In particular, for theories with $a>c$, increasing $g$ sufficiently, it is always possible to find entangling regions with arbitrarily negative values of the EE universal term.}. Now,  $\mathcal{W}_{\partial A} $ and $\mathcal{K}_{\partial A}$ are positive definite and positive semidefinite respectively, so it is straightforward to prove that a conjecture analogous to \req{ff0} in four dimensions, namely,
\begin{equation}\label{4df}
 \left. \frac{S^{4d}_{\rm log}(A)}{a}\right|_{\rm Maxwell} \leq  \frac{S^{4d}_{\rm log}(A)}{a} \leq   \left. \frac{S^{4d}_{\rm log}(A)}{a}\right|_{\rm free\, scalar} ,
\end{equation}
is trivially equivalent to the HM bounds of \req{HMbounds}. Since these have been rigorously proven in \cite{Hofman:2016awc}, \req{4df} is also true in general. In this case, the local nature of $S^{4d}_{\rm log}(A)$ limits the content of \req{4df} to be exactly equivalent to the one of the HM bounds. On the other hand, our three-dimensional conjecture (\ref{ff0}) contains much more information, as $F(A)$ does not have a closed geometric expression dependent on just a few coefficients which may be valid for general theories.

{\bf On the definition of $F(A)$:}
% obtain F0 from a direct calculation of
%9
%the entanglement entropy in a square lattice does not produce reasonable results.
%one obtains wildly oscillating answers as the ratio R/δ varies (here δ stands for the lattice spacing). The problem has to do with the fact that we cannot resolve the radius of the disk with a precision better than δ, which means that we cannot distinguish disks with radii R and R(1 + aδ), with a ∼ O(1). But such uncertainty will pollute F via the area-law piece
%in the entanglement entropy (1.1) as −F → −F + 2πc0a. As it is clear from this, the issue cannot be resolved by making the disk radius larger in the lattice.
Going back to three dimensions, let us start by observing that a direct computation of $F(A)$ from the EE formula using a lattice regularization does not produce unambiguous results in general. This has to do with the fact that it is not possible to resolve the characteristic scales of region $A$ with a precision better than the UV cutoff, \eg we cannot distinguish $R$ from $R(1+ a \delta)$ with $a \sim \mathcal{O}(1)$. This uncertainty pollutes $F(A)$ via the area-law term, $F\rightarrow F-a \cdot c_0\cdot \text{perimeter}(\partial A)$, and the situation cannot be improved by making $R$ larger in the lattice. 

In order to define $F(A)$ rigorously we can make use of mutual information \cite{Casini:2007dk,Casini:2008wt,Casini:2014yca,Casini:2015woa}. %, which is a well-defined quantity in the continuum \cite{}. 
 Given some region $A$ with characteristic scale $R$, consider two concentric regions $A^{-}$ and $\overline{A^{+}}$ with the same shape as $A$, defined by moving a distance $\varepsilon/2$ inwards and outwards, respectively, along the normal direction to $\partial A$. Then, the mutual information $I(A^+,A^-)$ tends, in the $\varepsilon/R \ll 1$ limit, to twice the EE of $A$, providing a well-defined notion of $F$ in the continuum, namely,
 \begin{equation}\label{muti}
 I(A^+,A^-)=\kappa \frac{\text{perimeter}(\partial A)}{\varepsilon}-2F(A)+\mathcal{O}(\varepsilon)\, .
 \end{equation} 
%Hence, ... \comment{local nature of slog implies only HM bound and nothing else; the sphere does not extremize EE across regions}
The robustness of this way of defining $F(A)$ has been previously exploited in several papers \cite{Casini:2015woa,Bueno:2021fxb,Huerta:2022cqw} and we will use it henceforth.

  {\bf  Orbifold theories and multicomponent regions: } 
Let us consider the case of orbifold theories $\mathrm{O}$ ---namely, theories obtained from the quotient of some complete parent theory $\mathrm{C}$ by some finite symmetry group $G$. For them, the mutual information is given by \cite{Casini:2019kex}
 \begin{align}\notag
   I^\mathrm{O}(A^+,A^-) &= I^\mathrm{C}(A^+,A^-)-n \log |G| \\ &+ \Delta \mathcal{S}(A^+)+\Delta \mathcal{S}(A^-)\, ,
  \end{align}
  where $n$ is the number of connected boundaries of $A$ and $|G|$ is the number of elements of $G$. For $A^\pm$ formed by more than one connected components  $A^\pm=A^\pm_1 \cup A^\pm_2 \cup \dots \cup A^\pm_k$, the quantities $\Delta \mathcal{S}(A^\pm)\equiv \mathcal{S}(\rho_{A^\pm} | \otimes_i^k \rho_{A_i^\pm})|_C-\mathcal{S}(\rho_{A^\pm} | \otimes_i^k \rho_{A_i^\pm})|_O $ are the differences of the relative entropies between the reduced density matrix on the region, and the tensor product of the density matrices reduced on each of its components. By monotonicity these 
differences are positive semi-definite.

   Hence, we can obtain $F(A)$ for a given orbifold theory in terms of the complete theory one using \req{muti}. One finds
  \begin{align}\notag
  \left. F(A)\right|_{\mathrm{O}} &= \left. F(A)\right|_{\mathrm{C}}+\frac{n}{2} \log |G| - \frac{1}{2}  [\Delta \mathcal{S}(A^+)+\Delta \mathcal{S}(A^-)]\, ,\\
   \left. F_0\right|_{\mathrm{O}} &= \left. F_0\right|_{\mathrm{C}}+\frac{1}{2} \log |G|\, .
  \label{111}
  \end{align}
   From this, it is easy to argue that orbifolding tends to decrease the value of $F(A)/F_0$, in agreement with our conjecture. %For instance, if the region has a single component with a single boundary, one easily finds   
%\begin{equation}
%1 \leq \left.  \frac{  F(A)}{F_0}\right|_{\mathrm{O}}=\frac{\left. F(A)\right|_{\mathrm{C}}+\frac{1}{2} \,\log |G|}{\left. F_0\right|_{\mathrm{C}}+\frac{1}{2}\, \log |G|}  \leq \left.  \frac{  F(A)}{F_0}\right|_{\mathrm{C}}  \, ,
%  \end{equation}
%Similarly, 
Indeed, consider a region with arbitrary topology.  In that case, we have 
  \begin{equation}\label{lowb}
 n\leq  \left.  \frac{  F(A)}{F_0}\right|_{\mathrm{O}}\le \frac{ \left.F(A)\right|_{\mathrm{C}}+\frac{n}{2}\, \log|G|}{\left. F_0\right|_{\mathrm{C}}+\frac{1}{2}\, \log |G|} \leq \left. \frac{  F(A)}{F_0}\right|_{\mathrm{C}} \, .
  \end{equation}
 % The first inequality can 
%  \comment{where does the first inequality come from? It would seem that one needs $(F(A)- n F_0)_{\mathrm C} \geq 1/2 [ [\Delta \mathcal{S}(A^+)+\Delta \mathcal{S}(A^-)]$ in order for it to hold}
The third inequality follows from $n\leq \left. (F(A)/F_0)\right|_{\mathrm{C}}$, proved in \cite{Bueno:2021fxb}, the second is a consequence of \req{111}, and the first follows from the semi-positivity of $\mathcal{S}(\rho_{A^\pm} | \otimes_i^k \rho_{A_i^\pm})|_O$. %, and the last one follows from the first, $n\leq \left. (F(A)/F_0)\right|_{\mathrm{C}}$ \cite{Bueno:2021fxb}. 
   Hence, the quotient for the parent theory is always greater than the one of the orbifold. Similarly, in all cases the lower bound is provided by the number of connected boundaries of the region.
Therefore, as far as our conjecture is concerned, any conclusions which hold for complete theories are also valid for orbifolds of such theories. 
 
 The same happens for infinite symmetry groups. In that case, $\log|G|$ is replaced by a divergent contribution, and the quotient saturates the lower bound appearing in \req{lowb}, namely,  $\left.(F(A)/F_0) \right|_{\mathrm{O}}=n$. This implies, in particular, that the Maxwell field, which is an orbifold of the free scalar theory  under the group $\mathbb{R}$ implementing $\phi\rightarrow \phi+ \lambda$ \footnote{It is equivalent to the theory of its derivatives, $\partial_{\mu}\phi = \varepsilon_{\mu\nu\delta} F^{\nu\delta}$} has
 \begin{equation}
 \left.\frac{F(A)}{F_0} \right|_{\rm Maxwell}= n \quad \forall \, A \, ,
 \label{333}\end{equation}
where $n$ is the number of connected boundaries of the region.  
More precisely, for the Maxwell field we get  $  \left. F(A)\right|_{\mathrm{Maxwell}}\sim \left. F(A)\right|_{\mathrm{free\, scalar}}+ n/4 \log (- \log (\delta))$ that diverges with the regularization scale $\delta$ \cite{Casini:2014aia}. 
The same saturation (\ref{333}) holds for topological models, for which $F(A)=\gamma \, n$ for some constant $\gamma$.
%As shown in \cite{Bueno:2021fxb}, the general inequality $1\leq F(A)/ F_0$ valid for general CFTs gets improved, in the case of regions with $n$ connected boundaries, to $n \leq F(A)/  F_0$. 
Hence, the lower bound in our general conjecture (\ref{ff0}) is not only consistent but fully equivalent to the general inequality $n \leq F(A)/  F_0$. % proved in  \cite{Bueno:2021fxb}. 
Let us now try to motivate the upper bound.
%we normalized the expression by the coefficient $a$, which is the one selected by a spherical entangling surface \cite{}, and theref.

{\bf Regions with disconnected components and large separations: }
In case there is a theory which provides an upper bound for $F(A)/F_0$ for general CFTs and arbitrary regions, this must be given by the free scalar. Indeed, consider an entangling region $A$ consisting of two disconnected subregions $A=A_1\cup A_2$. Then, we have $S(A_1\cup A_2) =S(A_1)+S(A_2)-I(A_1,A_2)$, where $I$ is the mutual information. Now, assume that $A_1$ and $A_2$ are both disk regions. Then, dividing both sides by $F_0$ and noticing that divergences cancel in both sides of the equality, one is left with 
$F(A_1\cup A_2)/F_0 = 2+I(A_1, A_2)/F_0$. Now, notice that in the long-distance regime the free scalar provides the greatest possible value of $I(A_1,A_2)$. Indeed, on general grounds, one has $I \sim \ell^{-4\Delta}$ where $\ell$ is the separation between regions and $\Delta$ is the smallest scaling dimension of the model. This is minimized by the free scalar in general dimensions, $\Delta_{\rm free\, scalar}=(d-2)/2$, which saturates the corresponding unitarity bound ---see \eg \cite{Minwalla:1997ka}. This means that $F(A)/F_0$ is absolutely maximized by the free scalar in that case. If one replaces now $A_1$ and $A_2$ by arbitrary shapes with characteristic lengths much smaller than $\ell$, the inequality \req{ff0} also holds provided it holds for $A_1$ and $A_2$ individually.

\begin{figure}[t] \centering
	\includegraphics[scale=0.67]{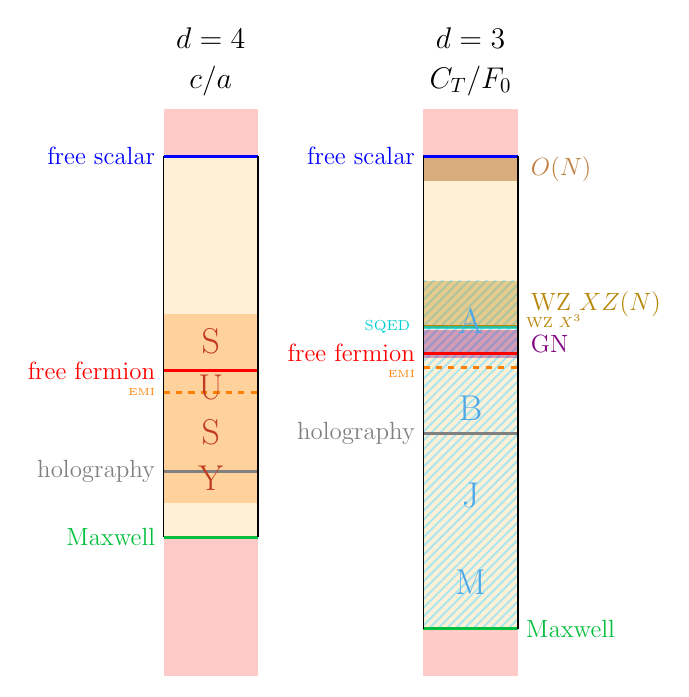} 	
	\caption{ \textsf{Quotients $c/a$ in $d=4$ and $C_{\ssc T}/F_0$ in $d=3$ for various CFTs. For visual clarity, each diagram is normalized by the free scalar result. For $d=4$, the unitarity bounds are known to be saturated by the free scalar and the Maxwell field, respectively. The holographic result, the free fermion and the EMI model (dashed orange) are also shown. For supersymmetric theories, the band of allowed values is smaller and appears displayed in pale orange. In $d=3$ the theories saturating the conjectural bounds are also the free scalar and the Maxwell field, for which $C_{\ssc T}/F_0|_{\rm Maxwell} =0$. Besides the free fermion, the EMI model and holography, we also present the range of values covered by various other theories: the $O(N)$ models for general $N$ (brown band), the Gross-Neveu models for general $N$ (purple band), the $\mathcal{N}=2$ Wess-Zumino model with superpotentials $X^3$ (orange line), $X \sum_i^N Z_iZ_i$ for general $N$ (pale brown band), SQED (green line) and general ABJM models (pale green band with diagonal lines). Red bands correspond to non-allowed values. } }
	\label{refiss376}
\end{figure}

{\bf Regions with disconnected components and thin deformations on a null cone:}
Additional evidence follows from the so-called {\it pinching property}. Consider the causal cone $H$ associated to some disk region $A$. Parametrizing the cone by some angular coordinate $\theta$ and an affine parameter $s\in (0,L)$, where $L$ is the position of $A$, the region $f(\delta,\epsilon)\equiv \{(\theta,s) / |\theta-\theta
_0| <\epsilon,s >\delta\}$  around some arbitrary direction $\theta_0$ is a sector of a conical  frustrum ---see Fig.\,4 in \cite{Casini:2021raa} for a drawing. Then, the region $A_1(\delta,\epsilon)\equiv H-f(\delta,\epsilon)$ corresponds to the causal cone of $A$ with such frustrum removed. $A_1$  has a boundary which corresponds to the original disk boundary for all $\theta$ except for $|\theta-\theta_0| <\epsilon$, in which case it is given by the boundary of $f$. Now, the pinching property establishes that, given some other arbitrary region $B$, \cite{Schlieder:1972qr,Wall:2011hj,Bousso:2014uxa,Casini:2021raa}
\begin{equation}
\begin{aligned}
\lim_{\epsilon\rightarrow 0}\lim_{\delta\rightarrow 0} I(A_1(\delta,\epsilon),B)&=0 \, , \,\, \text{(interacting CFT)} \\
\lim_{\epsilon\rightarrow 0}\lim_{\delta\rightarrow 0} I(A_1(\delta,\epsilon),B)&=I(A,B) \, , \,\, \text{(free CFT)}
\end{aligned}
\end{equation}
namely, for interacting CFTs ---including Generalized Free Fields \cite{Benedetti:2022aiw}--- the mutual information vanishes identically when we make the tip of $f$ approach the tip of the cone and then we make the conical sector arbitrarily thin. On the other hand, taking the same limits in the case of free CFTs ---in the sense of being fields satisfying a local  linear equation of motion--- we are instead left with the mutual information of the original disk region with $B$. Hence, considering the entanglement entropy for $A\equiv A_1(\delta,\epsilon)\cup A_2$ where $A_2$ is a disk region, we have
\begin{equation}
\begin{aligned}\label{fpinch}
 F(\lim_{\epsilon\rightarrow 0}\lim_{\delta\rightarrow 0} A)/F_0&=2 \, , \,\, \text{(interacting CFT)} \\ 
 F(\lim_{\epsilon\rightarrow 0}\lim_{\delta\rightarrow 0} A)/F_0&=2+\frac{I(A_1,A_2)}{F_0} \, , \,\, \text{(free CFT)}
\end{aligned}
\end{equation}
where by $A_1$ in the second line we mean the disk region which results from fully removing $f$. This holds regardless of the relative separation between $A_1$ and $A_2$. Hence, it is obvious that in this case $F(A)/F_0$ is smaller for any interacting CFT than for any free one. If the construction is repeated using regions other than disks, the same conclusion holds again as long as the individual regions satisfy \req{ff0}. Now, \req{fpinch} does not say anything about the hierarchy between the free theories themselves. However, strong numerical evidence suggests that   
\begin{equation}
   \left. \frac{I(A_1,A_2)}{F_0} \right|_{\rm free\,fermion} <  \left. \frac{I(A_1,A_2)}{F_0} \right|_{\rm free\, scalar}
 \end{equation}
  for arbitrary spatial regions $A_1,A_2$ \cite{Agon:2022efa}. Hence, once again we find that the free scalar provides an absolute maximum for $F(A)/F_0$ in this case.

{\bf Small deformations of a disk region:} Let us now consider regions with a single connected component.
 The first obvious case is the one of slightly deformed disks. We can parametrize their boundary by the radial equation 
\begin{equation}
\frac{r(\theta)}{R}=1+\frac{\epsilon}{\sqrt{\pi}} \sum_{\ell}  ( a_{\ell,(c)}\cos(\ell \theta) + a_{\ell,(s)} \sin(\ell \theta) )\, ,
\end{equation}
where $\epsilon \ll 1$. Then, at leading order in $\epsilon$, we have \cite{Allais:2014ata,Mezei:2014zla,Faulkner:2015csl}
\begin{equation}
\frac{F(A)}{F_0}=1+\frac{\pi^3 }{24} \frac{ C_{\ssc T}}{ F_0} \sum_{\ell} \ell (\ell^2-1) \left[ a_{\ell,(c)}^2 +a_{\ell,(s)}^2 \right] \epsilon^2\, ,
\end{equation}
where $C_{\ssc T}$ is the coefficient which controls, for a general CFT, the flat-space stress-tensor two-point function charge, namely,
\begin{equation}
\braket{T_{\mu\nu}(x) T_{\rho\sigma}(0)}_{\mathbb{R}^3}=\frac{C_{\ssc T}}{x^6} \left[I_{\mu(\rho} I_{\sigma) \nu}-\frac{\delta_{\mu\nu}\delta_{\rho\sigma}}{3} \right]\, ,
\end{equation}
where $I_{\mu\nu}\equiv \delta_{\mu\nu}-2 x_{\mu}x_{\nu} /x^2$. Now, noting that the coefficient which accompanies $\epsilon^2$ is positive semidefinite, applying our conjecture (\ref{ff0}) to the deformed-disks case we are left with a conjecture for the quotient of charges $C_{\ssc T}/F_0$, namely, with \req{ff04}.
 %\begin{equation}\label{ff04}
% 0\leq  \frac{C_{\ssc T}}{F_0} \leq \left. \frac{C_{\ssc T}}{F_0}\right|_{\rm free\, scalar}= \frac{3}{ 4\pi^2\log 2- 6\zeta[3]} \simeq  0.14887\dots % \left. \frac{F(A)}{F_0}\right|_{\rm Maxwell} 
%   \end{equation}
 %  for general three-dimensional CFTs
    In that expression, the lower bound becomes trivial, as  for the three-dimensional Maxwell field this quotient simply vanishes.
   As anticipated in the introduction, an inequality of this type is highly reminiscent of the four-dimensional HM bounds for the trace-anomaly coefficients $c/a$ ---see Fig.\,\ref{refiss376}. 
   
   As it turns out, both $C_{\ssc T}$ and $F_0$ have been computed for a plethora of three-dimensional CFTs and we can test the validity of \req{ff04} in all those cases. In the appendix we have gathered the results, and in Fig.\,\ref{refiss376} we have plotted them together. We observe that all considered theories satisfy the conjectural bounds. In particular, one finds a similar hierarchy as in the four-dimensional $c/a$ case, with the free scalar \cite{Osborn:1993cr,Marino:2011nm,Klebanov:2011gs,Casini:2009sr} representing the upper bound, the free fermion \cite{Osborn:1993cr,Klebanov:2011gs,Marino:2011nm,Casini:2009sr} taking a lower value, holographic theories  \cite{Liu:1998bu,Ryu:2006ef} an even lower one, and the Maxwell field providing the lowest possible one (zero in the three-dimensional case). Explicitly, we have: $\left. (C_{\ssc T}/F_0) \right|_{\text{free scalar}}=3/(4\pi^2\log 2- 6\zeta(3)) \simeq 0.14887$,  $\left. (C_{\ssc T}/F_0) \right|_{\text{free fermion}}=3/(4\pi^2\log 2+ 6\zeta(3)) \simeq 0.086764$, $\left. (C_{\ssc T}/F_0) \right|_{\text{EMI}}=8/\pi^4 \simeq 0.082128$,  $\left. (C_{\ssc T}/F_0) \right|_{\text{holography}}=6/\pi^4 \simeq 0.061596 $, $\left. (C_{\ssc T}/F_0) \right|_{\text{Maxwell}}= 0$,
 %  \begin{align}
%& \left.  \frac{ C_{\ssc T}}{F_0}\right|_{\text{free scalar}}&&= \frac{3}{ 4\pi^2\log 2- 6\zeta(3)} \simeq 0.14887 \, ,\\
%&  \left.  \frac{ C_{\ssc T}}{F_0}\right|_{\text{free fermion}}&&=   \frac{3}{ 4\pi^2\log 2+6\zeta(3)} \simeq 0.086764 \, ,\\
% &   \left.  \frac{ C_{\ssc T}}{F_0}\right|_{\text{EMI}}&&= \frac{8}{\pi^4} \simeq 0.082128 \, ,\\
%   &     \left.  \frac{ C_{\ssc T}}{F_0}\right|_{\text{holography}}&&= \frac{6}{\pi^4} \simeq 0.061596 \, ,\\
%    &           \left.  \frac{ C_{\ssc T}}{F_0}\right|_{\text{Maxwell}}&&=0 \, .
 %  \end{align}
   where we have also included the result corresponding to the so-called ``Extensive Mutual Information'' (EMI) model \cite{Casini:2008wt,Agon:2021zvp}.
   
 Amongst the interacting theories considered, we have the Gross-Neveu $O(N)$ UV fixed points models \cite{Klebanov:2011gs,Giombi:2014xxa,Diab:2016spb}, for which it is possible to find values  of $N$ which are both greater and smaller than the free fermion one, the whole range being $0.0854 \lesssim \left. C_{\ssc T}/F_0\right|_{{\rm GN},\,  O(N)} \lesssim 0.094$ $\forall\, N$. On the other hand, for the Wilson-Fisher fixed points of the scalar $O(N)$ models \cite{Klebanov:2011gs,Giombi:2014xxa,Petkou:1994ad,ElShowk:2012ht,Kos:2013tga}, one finds that the free scalar result is always an upper bound, for arbitrary values of $N$. The range is  $0.1409\lesssim   \left. C_{\ssc T}/F_0\right|_{ O(N)} \leq \left. C_{\ssc T}/F_0\right|_{ \text{free scalar}} $ $\forall\, N$.  Additional theories considered include various supersymmetric $\mathcal{N}=2$ Wess-Zumino models as well as general $U(N)_k\times U(N)_{-k}$ ABJM models \cite{Aharony:2008ug}, for which we find ---using results from \cite{Kapustin:2010xq,Marino:2011eh,Nishioka:2013haa,Chester:2014fya,Binder:2020ckj,Alday:2021ymb,Codesido:2014oua,Agmon:2017xes,Bobev:2022eus}--- that $0\leq \left. C_{\ssc T}/F_0\right|_{U(N)_k\times U(N)_{-k}\, {\text{ABJM}}} \leq 3/(2\pi^2 \log 4)\simeq 0.10963$ for all $N$ and $k$ ---see Fig.\,\ref{fig:ABJM} in the appendix. In all cases, the conjectural bounds are respected. It would be certainly interesting to test the conjecture for additional theories.

\begin{figure}[t] \centering
	\includegraphics[width=0.49\textwidth]{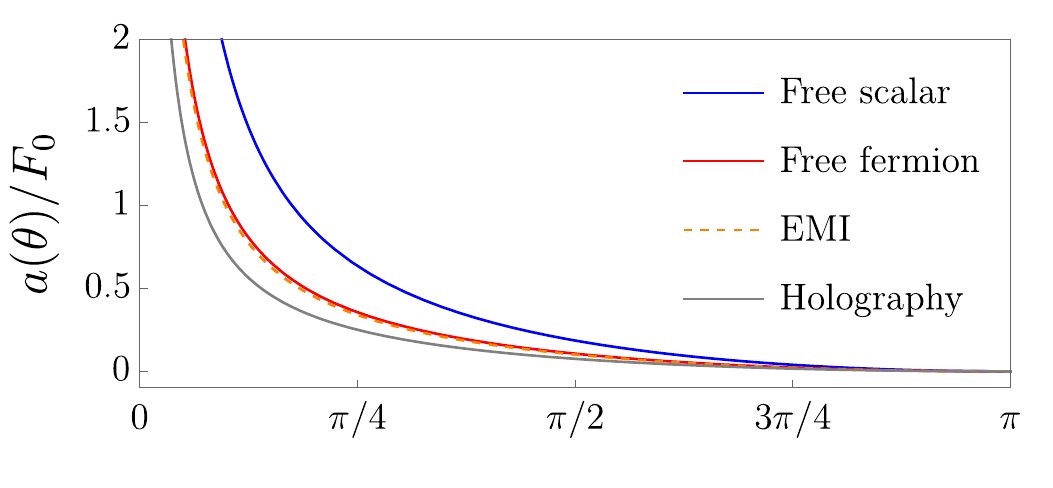}

\vspace{-0.5cm}	
	
	\includegraphics[width=0.49\textwidth]{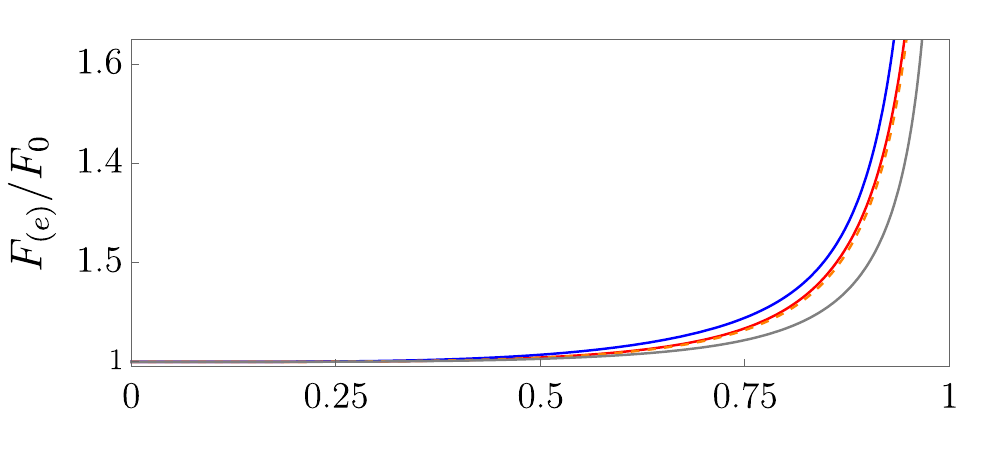}  	
	\caption{ \textsf{We plot the EE universal coefficients corresponding, respectively, to a corner region of opening angle $\theta$, and an ellipse of eccentricity $e$ ---both normalized by $F_0$--- as functions of those parameters for: a free scalar (blue), a free fermion (red), the EMI model (dashed orange) and holographic Einstein gravity (gray). In all cases, the free scalar one lies above the curves of all the rest of theories. The Maxwell field is a trivial lower bound of constant value ($0$ and $1$, respectively) in both cases. } }
	\label{refiss3746}
\end{figure}

{\bf Ellipses and corners:} Moving from the perturbed-disks regime, values of $F(A)/F_0$ for more complicated regions exist in some cases, at least for a few theories. In particular, there exist results for free scalars and fermions, the EMI model, as well as for holographic theories in the case of ellipses of arbitrary eccentricity \cite{Bueno:2021fxb}\footnote{The results displayed in Fig.\,\ref{refiss3746} are obtained using a combination of analytic methods (in the limiting cases of very thin and almost round ellipses), lattice calculations, and certain trial functions defined in \cite{Bueno:2021fxb} which approximate the exact curves. In all cases, the small uncertainties in the approximations will not alter the hierarchy of theories shown in the figure.}. The results are shown in Fig.\,\ref{refiss3746}, where it is clear that the free scalar always takes the greatest value. The lower bound provided by the Maxwell theory is always trivially satisfied by all theories since one has $\left. (F_{(e)}/F_0)\right|_{\rm Maxwell}=1\, \forall\,  e$.

 In Fig.\,\ref{refiss3746} we have also presented results for the same set of theories in the case of corner regions of opening angle $\theta$ \cite{Casini:2006hu,Casini:2008as,Casini:2009sr,Bueno1,Bueno3,Elvang:2015jpa,Hirata:2006jx,Helmes:2016fcp}. In that case, $F(A)$ builds up a logarithmic divergence weighted by some function $a(\theta)$ which, normalized by $F_0$, inherits  the same hierarchies as in \req{ff0}. Again, the free scalar curve ---which also coincides with the one of the large-$N$ limit of the Wilson-Fisher $O(N)$ model \cite{Whitsitt:2016irx}--- is above all others. On the other hand, there exists a general lower bound for $a(\theta)$ constructed in \cite{Bueno:2015ofa} and given by $a(\theta)/F_0 \geq \frac{\pi^2 C_{\ssc T}}{3F_0}  \log[1/\sin(\theta/2)]$. In the case of the Maxwell field, the right-hand-side is just zero, so again we find consistency with the lower bound in \req{ff0}. Computations of the corner function with $\theta=\pi/2$ have been performed using numerical methods for the $O(N)$ models with $N=1,2,3$ \cite{Tagliacozzo:2009jca,2013PhRvL.110m5702K,Stoudenmire:2014hja,Kallin:2014oka,Sahoo:2015hma}. In all cases, the result is very close to the free scalar one, but the precision achieved does not seem to allow for a trustworthy quantitative comparison \footnote{The values provided \eg in \cite{2013PhRvL.110m5702K,Stoudenmire:2014hja,Kallin:2014oka} would seem to suggest that $a(\pi/2)/F_0$ could be greater than the corresponding free scalar result for these theories. However, as shown in Fig.\,5(a) of \cite{Sahoo:2015hma}, increasing the precision of the Numerical Linked Cluster Expansion (NLCE) tends to decrease the value of $a(\pi/2)$. A naive extrapolation of the NCLE order obtained from the values presented in that paper for the Ising model yields a value of $a(\pi/2)/F_0$ comfortably smaller than the free scalar one. Such value is similar to the one obtained in \cite{Tagliacozzo:2009jca} using a tensor network variational ansatz, namely, $\left.(a(\pi/2)/F_0)\right|_{\rm Ising} \approx 0.16 \approx 0.85 \left.(a(\pi/2)/F_0)\right|_{\rm free\, scalar} $ }.
%The quantity $F$ defined through the mutual information does not contain topological contributions, we take this notion. Also, argue that orbifold theories do not count as complete theories (otherwise one could violate the bounds). \comment{autorecurrente a traves de la mutual}
%\\

Both for ellipses and corners, the behavior in the regime in which the region becomes very sharp ---\ie for $e,\theta \rightarrow 0$, respectively--- is controlled by the universal coefficient characterizing the EE of a thin strip. Given such a strip of dimensions $L,r$ with $L\gg r$, one finds
\begin{equation}\label{conjk}
\frac{F(A)}{F_0}=\frac{\kappa}{F_0}\frac{L}{r}+ \mathcal{O}(r^0)\, .  %\quad \mathcal{R}=\frac{L}{\pi r} \quad \Rightarrow \quad \frac{\kappa}{F_0} \overset{?}{\geq} \frac{1}{\pi}\simeq 0.31831\, .
\end{equation}
The coefficient $\kappa$ is yet another quantity characterizing any given three-dimensional CFT. It is not known to be related with any other coefficient defined beyond EE, so our general conjecture (\ref{ff0}) predicts additional independent bounds on the possible values of $\kappa/F_0$. Using the free scalar values of $\kappa$ computed in \cite{Casini:2009sr}, one finds
\begin{equation}\label{boundik}
 0 \leq  \frac{\kappa}{F_0}  \leq   \left. \frac{\kappa}{F_0} \right|_{\rm free\, scalar}\simeq 0.6223\dots
\end{equation}
The values of $\kappa$ are also known for free fermions \cite{Casini:2009sr}, the EMI model \cite{Casini:2008wt}, as well as for holographic theories dual to Einstein gravity \cite{Ryu:2006bv}. In each of those cases, one finds $\left. (\kappa/F_0) \right|_{\text{free fermion}} \simeq 0.3297$, $\left.  (\kappa/F_0) \right|_{\text{EMI}}= 1/\pi \simeq 0.3183 $,  $\left.  (\kappa/F_0)  \right|_{\text{holography}}=2\Gamma[\tfrac{3}{4}]^2/\Gamma[\tfrac{1}{4}]^2\simeq 0.2285 $, always in agreement with \req{boundik}. Naturally, using \req{boundik} we can obtain putative bounds for $\kappa$ for any CFT for which $F_0$ is known.  %For instance, we have $\left.  \kappa \right|_{\text{Ising}}\lesssim0.0380  $ or  $\left.  \kappa \right|_{U(1)_k\times U(1)_{-k}\, \text{ABJM} }\lesssim  0.6223\log(4k)$ for the Ising and $N=1$ ABJM models, respectively. 
Evaluating this coefficient for %these or 
additional theories would be another way of testing our general conjecture.

{\bf Discussion:} In this letter we have presented evidence in favor of  a new conjecture for the EE universal coefficient of general three-dimensional CFTs. As we have seen, the conjecture fits very well with previous results like the HM bounds in $d=4$ as well as with the fact that $F(A)/F_0$ is bounded below by the number of connected boundaries of $A$ for general theories. Naturally, it would be very interesting to find a proof (or a counterexample) to our conjectures. This would entail a better understanding of what makes the free scalar theory special from an entropic point view.

An obvious question is whether our conjecture may  also extend to higher dimensions. In $d=5$, an analogous putative upper bound corresponding to a free scalar would imply ---via the perturbed spheres EE \cite{Mezei:2014zla},
\begin{equation}\label{cinc}
%0= \left.\frac{C_{\ssc T}}{F_0}\right|^{d=5}_{\rm Maxwell} \leq
\left. \frac{C_{\ssc T}}{F_0}\right|^{d=5} \leq \left.\frac{C_{\ssc T}}{F_0}\right|^{d=5}_{\rm free\, scalar}%=\frac{45}{2\pi ^4 \log 2+2 \pi ^2 \zeta (3)-15 \zeta (5)}
\simeq 0.314221\dots
\end{equation} %\vspace{0.1cm}
%
%\noindent
The analogous bound on the strip coefficient would be
\begin{equation}\label{cinc2}
%0= \left.\frac{\kappa}{F_0}\right|^{d=5}_{\rm Maxwell} \leq
 \left.\frac{\kappa}{F_0} \right|^{d=5}\leq \left.\frac{\kappa}{F_0}\right|^{d=5}_{\rm free\, scalar}%=\frac{45}{2\pi ^4 \log 2+2 \pi ^2 \zeta (3)-15 \zeta (5)}
\simeq 0.228104\dots
\end{equation}
In both cases, the lower bound provided by the Maxwell field would always be trivially satisfied, since $F_0$ diverges for that theory \cite{Giombi:2015haa}.
 It is easy to check that both \req{cinc} and \req{cinc2} are satisfied for free fermions as well as for holographic theories. A related question is whether or not the round $\mathbb{S}^3$ is the entangling surface with the smallest $F(A)$  in $d=5$. %, which again would motivate using that value as a normalization. 
 A study of the $d=6$ case would also be interesting. This would be trickier than in $d=4$ since there are four trace-anomaly coefficients rather than two, and the geometric expression of $S^{6d}_{\rm log}(A)$ is considerably more involved \cite{Safdi:2012sn,Miao:2015iba}.

\vspace{0.3cm}
{\it Acknowledgments:} We thank Oren Bergman, Nikolay Bobev, Alba Grassi, Niko Jokela, Kyriakos Papadodimas and Kazuya Yonekura for useful discussions. The work of PB was supported by a Ram\'on y Cajal fellowship (RYC2020-028756-I) from Spain’s Ministry of Science and Innovation. The work of HC was supported by the Simons foundation through the It From Qubit Simons collaboration and by CONICET, CNEA, and Universidad Nacional de Cuyo, Argentina. The work of JM is partially supported by the Israel Science Foundation, grant no. 1487/21.

\vspace{-1cm} 
\onecolumngrid % \vspace{1cm} 
%\begin{center}  
%{\Large\bf Appendices} 
%\end{center} 
\appendix 
%\tableofcontents  

\section{$C_{\ssc T}/F_0$ for various three-dimensional CFTs}
In this appendix we present the values of $C_{\ssc T}/F_0$  for a collection of three-dimensional CFTs. In particular, we show results for free scalars and fermions, the EMI model, theories with a holographic Einstein dual, the Wilson-Fisher fixed points of the $O(N)$ models, the UV fixed points of the Gross-Neveu models, various supersymmetric Wess-Zumino models with different superpotentials and $U(N)_k\times U(N)_{-k}$ ABJM models for general values of $N$ and $k$. We also comment on the general form of the quotient $C_{\ssc T}/F_0$ for holographic higher-curvature gravities. 

\subsection{Free theories}
The exact values of
$C_{\ssc T}$ and $F_0$  for free scalars and fermions are given by \cite{Osborn:1993cr,Marino:2011nm,Klebanov:2011gs,Casini:2009sr}
\begin{align}
\left.C_{\ssc T}\right|_{\rm free\, scalar} &=\frac{3}{32\pi^2} \, ,  \quad \left.F_{0}\right|_{\rm free\, scalar}=\frac{1}{16}  \left[2\log 2 - \frac{3}{\pi^2} \zeta(3) \right]\, , \\
\left.C_{\ssc T}\right|_{\rm  free\, fermion}&=\frac{3}{16\pi^2} \, , \quad \left.F_{0}\right|_{\rm free\,  fermion}=\frac{1}{8} \left[2\log 2 + \frac{3}{\pi^2} \zeta(3) \right] \, , 
\end{align}
so the ratios are
\begin{equation}
\left.\frac{C_{\ssc T}}{F_{0}}\right|_{\rm free\, scalar} = \frac{3}{ 4\pi^2\log 2- 6\zeta(3)} \simeq 0.148869\dots \, , \quad  \left.\frac{C_{\ssc T}}{F_{0}}\right|_{\rm free\, fermion}=   \frac{3}{ 4\pi^2\log 2+6\zeta(3)} \simeq 0.0867636\dots
\end{equation}

\subsection{Extensive mutual information model}
The ``Extensive mutual information model'' \cite{Casini:2008wt} follows from considering a general expression for the mutual information that satisfies all the known general axioms for this quantity in a general QFT \cite{Agon:2021zvp}. In additional to these, it satisfies the condition of being an extensive function of its arguments. This is equivalent to impose that the tripartite information vanishes for any regions $A$, $B$ and $C$, this is, $I_3\left(A,B,C\right)\equiv I\left(A,B\right)+I\left(A,C\right)-I\left(A,BC\right)=0$. The model is equivalent to a free fermion in $d=2$ \cite{Casini:2005rm} but the identification is lost for higher dimensions \cite{Agon:2021zvp}. This can be easily checked from the inequivalence of the quotients $c/a$ and $C_{\ssc T}/F_0$ displayed in Figure \ref{refiss376}. For the latter, the numerical value is given by
\begin{equation}
\left.\frac{C_{\ssc T}}{F_0}\right|_{\rm EMI}=\frac{8}{\pi^4} \simeq 0.0821279\dots
\end{equation}
In spite of differing from an actual theory for $d\geq3$, the EMI model still satisfies all known requirements for a valid mutual information, providing an useful toy model for many purposes ---see \eg \cite{Casini:2015woa,Bueno1,Witczak-Krempa:2016jhc,Bueno:2019mex,Estienne:2021lxh,Bueno:2021fxb}. %In particular, it was recently shown that the entanglement entropy of regions with complicated geometries was derived from simple expressions \cite{Bueno:2021fxb}.

%$/=8/\pi^4\simeq 0.08213 $, 

\subsection{Theories with a holographic Einstein dual}
For holographic theories dual to Einstein gravity in the bulk, one finds
\cite{Liu:1998bu,Ryu:2006ef}
\begin{align}\label{holoE}
\left.C_{\ssc T}\right|_{\rm holography}&=\frac{3}{\pi^3} \frac{L_{\star}^2}{G} \, ,  \quad \left.F_{0}\right|_{\rm  holography}=\frac{\pi}{2} \frac{L_{\star}^2}{ G} \, , \quad \text{so} \quad \left.\frac{C_{\ssc T}}{ F_{0}}\right|_{\rm holography}=\frac{6}{\pi^4} \simeq 0.0615959\dots
\end{align}
where $L_{\star}$ is the AdS radius and $G$ is the Newton constant. This result is reproduced by supersymmetric gauge theories with a holographic dual in the large $N$ limit. %This includes various theories such as the ABJM models.

%\subsection{Large $N$ superconformal gauge theories }
% In particular, in the large-$N$ limit of certain $\mathcal{N}=2$ supersymmetric gauge theories one finds \cite{} \comment{also for $\mathcal{N}=1$, I think}
%\begin{equation}
%\frac{C_{\ssc T}^{{\rm SUSY\, gauge},\, {\rm large\,}N}}{F_0^{{\rm SUSY\, gauge},\, {\rm large\,}N}}=\frac{6}{\pi^4 }\, ,
%\end{equation}
%which coincides with the Einstein gravity result and also saturates the lower bound.  

\subsection{  $O(N)$  models }
There are also available results for the Wilson-Fisher fixed points of the $O(N)$ model both for large $N$ and for small values of $N$. We have for the first two orders in the large-$N$ expansion \cite{Petkou:1994ad,Klebanov:2011gs}
\begin{align}
&\left.C_{\ssc T}\right|_{O(N)}= N \left.C_{\ssc T}\right|_{\rm free\, scalar} \left[ 1-\frac{40}{9\pi^2}\frac{1}{N} \right] +\mathcal{O}(1/N)\, , \\
&\left.F_{0}\right|_{O(N)}=N  \left.F_{0}\right|_{\rm free\, scalar} \left[1-\frac{2\zeta(3)}{\pi^2  \left[2 \log 2 -\frac{3}{\pi^2}\zeta(3) \right]}\frac{1}{N} \right]+\mathcal{O}(1/N)\, , 
\end{align}
and from this
\begin{align}
\left.\frac{C_{\ssc T}}{F_{0}}\right|_{O(N)} &= \left.\frac{C_{\ssc T}}{F_0}\right|_{\rm free\, scalar} \left[1- \left(\frac{40}{9\pi^2}-\frac{2\zeta(3)}{\pi^2 \log 4- 3\zeta(3)} \right) \frac{1}{N} \right]+\mathcal{O}(1/N^2) \\ \notag & \simeq \left.\frac{C_{\ssc T}}{F_0}\right|_{\rm free\, scalar} \left[1- \frac{0.21172}{N} \right]+\mathcal{O}(1/N^2)\, .
\end{align}
Hence, the result falls inside the window allowed by the conjectural bounds. For finite values of $N$, the result becomes smaller and tends to deviate from the free scalar result.  For instance, for $N=20$ one finds $C_{\ssc T}^{O(20)}/F_{0}^{O(20)}\simeq 0.1487$. %and for $N=5$,  $C_{\ssc T}^{O(5)}/F_{0}^{O(5)}\simeq 0.1426$.
There are also available results for lower values of $N$. For the Ising model, corresponding to $N=1$, one has \cite{ElShowk:2012ht,Kos:2013tga,Giombi:2014xxa} 
\begin{equation}
\left.C_{\ssc T}\right|_{\rm Ising}\simeq 0.9466   \left.C_{\ssc T}\right|_{\rm free\, scalar}\, , \quad \left.F_{0}\right|_{\rm Ising}\simeq 0.957  \left.F_{0}\right|_{\rm free\, scalar}\, , \quad \text{so }\quad \left.\frac{C_{\ssc T}}{F_{0}}\right|_{\rm Ising}\simeq  0.1468\, ,
\end{equation}
which lies again within the range. Similarly, for $N=2$ and $N=3$ one finds \cite{Kos:2013tga,Giombi:2014xxa}
\begin{equation}
\left.\frac{C_{\ssc T}}{F_{0}}\right|_{O(2)}\simeq 0.1469\, , \quad \left.\frac{C_{\ssc T}}{F_{0}}\right|_{O(3)}\simeq 0.147\, ,
\end{equation}
again within the range.

\subsection{ Gross-Neveu models }
There are also results for the UV fixed point of the Gross-Neveu models. In particular, in the large limit we have \cite{Klebanov:2011gs,Diab:2016spb}
\begin{align}
&\cTb{{\rm GN},\,  O(N)}= N \cTb{\rm free\, fermion} \left[ 1+\frac{8}{9\pi^2}\frac{1}{N} \right] +\mathcal{O}(1/N)\, , 
\\
&\Fb{{\rm GN},\,  O(N)}=N  \Fb{\rm free\,  fermion} \left[1+\frac{2\zeta(3)}{\pi^2  \left[2 \log 2 +\frac{3}{\pi^2}\zeta(3) \right]}\frac{1}{N} \right]+\mathcal{O}(1/N)\, , 
\end{align}
and hence
\begin{align}
\cTFb{{\rm GN},\,  O(N)}&= \cTFb{\rm free\, fermion} \left[1+ \left(\frac{8}{9\pi^2}-\frac{2\zeta(3)}{\pi^2 \log 4+ 3\zeta(3)} \right) \frac{1}{N} \right]+\mathcal{O}(1/N^2) \\ & \notag \simeq \cTFb{\rm free\, fermion} \left[1- \frac{0.048997}{N} \right]+\mathcal{O}(1/N^2)\, ,
\end{align}
which is also comfortably within the range. In order to probe small values of $N$, we can use the approximation found in \cite{Giombi:2014xxa} 
\begin{equation}
\Fb{{\rm GN},\,  O(N)}\simeq 0.103154 N +0.0516955 -\frac{\pi N}{96(N+6)}
\end{equation}
and the results obtained in \cite{Diab:2016spb} for $\cTb{{\rm GN},\,  O(N)}$. One finds, for instance
 \begin{equation}
\cTFb{{\rm GN},\,  O(4)}\simeq 0.08544 \, , \quad \cTFb{{\rm GN},\,  O(8)}\simeq 0.0895\, ,
\end{equation}
which once more lie within the range. The values tend to grow slightly as $N$ increases, but they stop doing so at some point ---the maximum seems to be (conservatively) lower than $0.094$. For instance, for $N=100$ one finds, from the large $N$ approximation,
 \begin{equation}
\cTFb{{\rm GN},\,  O(100)}\simeq 0.08672\, .
\end{equation}
Observe that as we vary $N$, in some cases the value  of the ratio is greater than the free fermion one, and in others it is smaller. This is different from the situation encountered for the $O(N)$ models, in which case the ratio is smaller than the free scalar one $\forall \, N$.

%It is interesting that, as opposed to the $O(N)$ models, which never surpass the free scalar value for the ratio

\subsection{$\mathcal{N}=2$ Wess-Zumino models}
The free energy and central charge corresponding to the critical Wess-Zumino model, also known as supersymmetric Ising model \cite{Bobev:2015vsa}, ---which has a cubic superpotential $\mathcal{W}=X^3$--- are given by \cite{Nishioka:2013gza,Giombi:2014xxa}
\begin{equation}
\left.C_{\ssc T}\right|_{X^3}\simeq 0.02766 \, , \quad \left.F_0\right|_{X^3}\simeq 0.29079 \, ,\quad \text{so} \quad
\left.\frac{C_{\ssc T}}{F_0}\right|_{X^3}\simeq 0.095122\, ,
\end{equation}
which falls within the range.  
%\begin{equation}
%F^{\rm }
%\end{equation}
%with $R$-charge $\Delta$, are given by \cite{Nishioka:2013gza}
%\begin{align}
%F^{{\rm WZ} , \Delta}_0 &=-\int_0^{\infty} \frac{\diff x}{2x} \left[\frac{\sinh (2(1-\Delta)x)}{\sinh^2 x} - \frac{2(1-\Delta)}{x} \right]\, , \\
%C_{\ssc T}^{{\rm WZ}, \Delta} &=  \int_0^{\infty}  \frac{3\diff x }{\pi^4} \left[ (1-\Delta) \left[\frac{1}{x^2}-\frac{\cosh(2x(1-\Delta))}{\sinh^2 x} \right]+\frac{[\sinh(2x)-2x] \sinh(2x(1-\Delta))}{2\sinh^4 x} \right] \, .
%\end{align}
%This contains information about various CFTs. On the one hand, $\Delta=1/2$ corresponds to 
%For all allowed values of $\Delta \in [1/2,1)$ we find that both bounds are satisfied. In particular, we have
%$
%C_{\ssc T}^{{\rm WZ}, \Delta} /F^{{\rm WZ} ,  \Delta}_0\simeq 0.1096, 0.1004, 0.09496, 0.09160, 0.08955,  0.08843 \, .
%$ for $\Delta=6/12,7/12,\dots,11/12$ respectively.

On the other hand, for the critical point of a Wess-Zumino model with superpotential $\mathcal{W}=X \sum_{i=1}^N Z^i$  we have, in the large $N$ limit  \cite{Nishioka:2013gza}
\begin{equation}
\left.C_{\ssc T}\right|_{XZ(N)} = \frac{3N}{8\pi^2}-\frac{2}{\pi^4}+ \frac{1}{N} \left(\frac{68}{9\pi^2}-\frac{48}{\pi^4} \right)+\mathcal{O}(1/N^2)\, , \quad  \left.F_0\right|_{XZ(N)}=\frac{N}{2}\log 2 + \frac{4}{\pi^2 N}+ \mathcal{O}(1/N^2)\, ,
\end{equation}
so
\begin{equation}
\left.\frac{C_{\ssc T}}{F_0}\right|_{XZ(N)}\simeq 0.10963- \frac{0.05924}{N}+\mathcal{O}(1/N^2)\, .
\end{equation}
For small values of $N$, on the other hand, we have \cite{Nishioka:2013gza}
\begin{align}\notag
&\left.\frac{C_{\ssc T}}{F_0}\right|_{ XZ(1)}\simeq 0.09706\, , \quad \left.\frac{C_{\ssc T}}{F_0}\right|_{XZ(2)}\simeq  0.09499\, , \quad \left.\frac{C_{\ssc T}}{F_0}\right|_{ XZ(3)}\simeq 0.09593\, , \quad \left.\frac{C_{\ssc T}}{F_0}\right|_{\rm XZ(4)}\simeq 0.09755\,, \\ & \notag \left.\frac{C_{\ssc T}}{F_0}\right|_{XZ(5)}\simeq  0.09929\,, \quad \cTFb{XZ(6)}\simeq 0.10059 \, , \quad  \cTFb{ XZ(7)}\simeq 0.10163 \, , \quad 
\cTFb{XZ(8)}\simeq 0.10250 \, ,\\ &  \cTFb{XZ(9)}\simeq 0.10326 \, , \quad \cTFb{XZ(10)}\simeq 0.10385\,. \quad 
\end{align}
As we can see, in all cases the values fall within the range. The case $N=2$ is particularly interesting, as it corresponds also to the $XYZ$ model \cite{Giombi:2014xxa} ---defined by a superpotential $\mathcal{W}=XYZ$--- which is in turn mirror symmetric to $\mathcal{N}=2$ SQED \cite{Aharony:1997bx,deBoer:1997ka}. For the latter two models, we have \cite{Jafferis:2010un,Bobev:2015jxa,Witczak-Krempa:2015jca,Jian:2016zll}
\begin{align}
&\Fb{XYZ}=\Fb{{\mathcal{N}}=2\, \rm SQED}=-3\ell(1/3) \simeq 0.87237\, , \\
&\cTb{XYZ}=\cTb{{\mathcal{N}}=2\, \rm SQED}=\frac{2}{27\pi^2} \left[16- \frac{9\sqrt{3}}{\pi} \right]\simeq 0.082844
\end{align}
where we used the definition \cite{Jafferis:2010un}
\begin{equation}
\ell(z)\equiv -z \log \left(1-\expp{2\pi \iu z} \right)+\frac{\iu}{2} \left(\pi z^2+\frac{1}{\pi} {\rm Li}_2(\expp{2\pi \iu z}) \right) -\frac{\iu\pi}{12}\, .
\end{equation}
From this, one finds
\begin{equation}
\cTFb{XYZ}= \cTFb{{\mathcal{N}}=2\, \rm SQED}\simeq 0.09492\, , 
\end{equation}
which agrees well with the result from \cite{Nishioka:2013gza} for the $X Z(2)$ model presented above.

 We also have results for the fixed point corresponding to a superpotential $\mathcal{W}=\sum_{i=1}^N (Z_iZ_i)^2$ \cite{Nishioka:2013gza}
\begin{equation}
\cTb{Z^4(N)}= \frac{3N}{8\pi^2}\, , \quad \Fb{{ Z^4(N)}}= \frac{N}{2}\log 2 \, ,\quad \text{so} \quad
\cTFb{Z^4(N)}=\frac{3}{4\pi^2 \log 2}\simeq 0.10963\, .
\end{equation}
The results are actually identical to the ones obtained for $N$ free chiral multiplets and also satisfy the bounds.

\subsection{$U(N)_k\times U(N)_{-k}$ ABJM model}

\begin{figure}[t] \centering
	\includegraphics[width=0.75\textwidth]{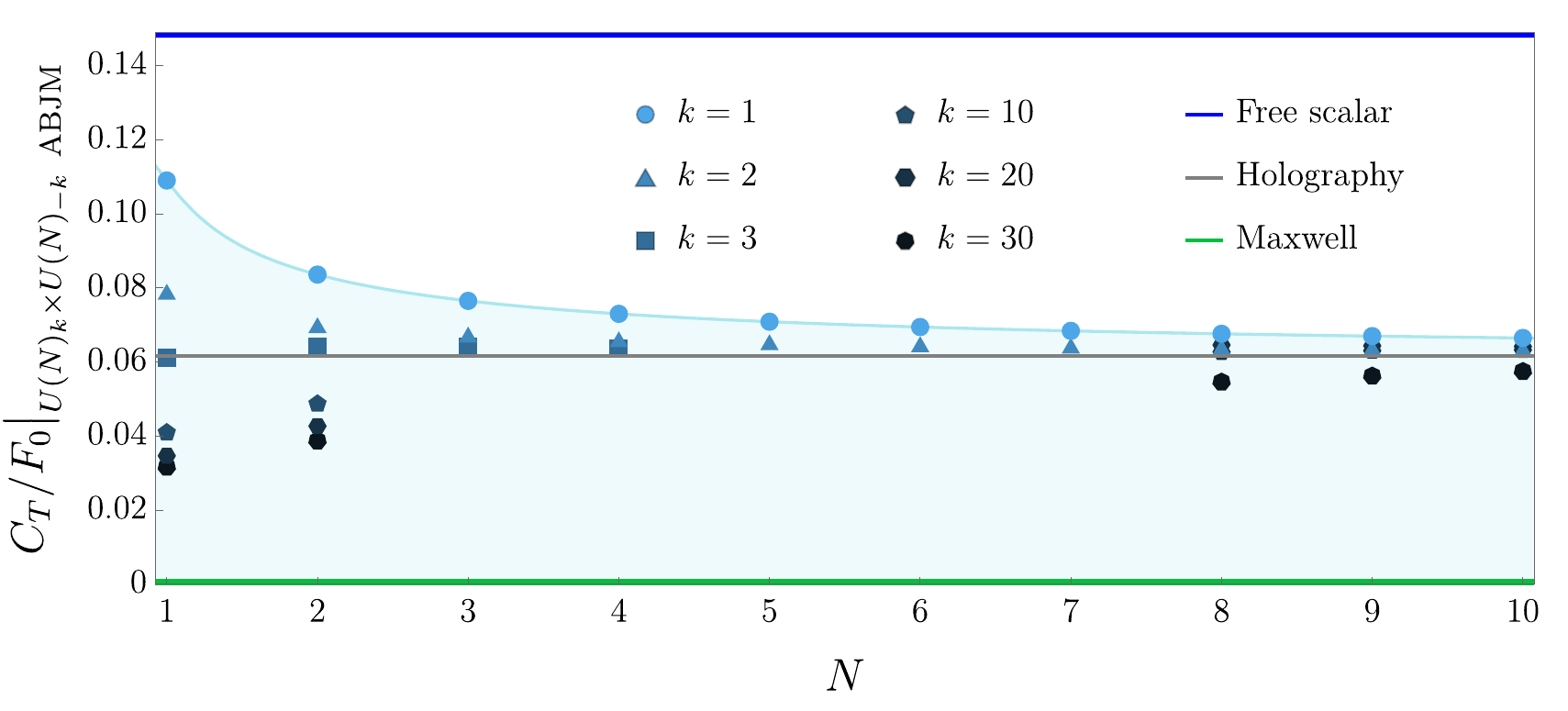}
\caption{Values of $C_{\ssc T}/F_0\big|_{U(N)_k\times U(N)_{-k}\text{ ABJM}}$ for several values of $N$ and $k$, namely: i) for $k=1$ and $k=2$ we represent the quotients for $N=1,\ldots ,10$ as circles and triangles respectively. ii) for $k=3$ we include the quotients for $N=1,2,3$ and $4$ with squares. iii) For $k=10, 20$ and $30$ we plot the values $N=1,2,8,9,10$, represented by pentagons, hexagons and heptagons respectively. All values we explored are contained in the blue region, whose lower bound is given by $C_{\ssc T}/F_0\big|_{U(N)_k\times U(N)_{-k}\text{ ABJM}}\geq0$ ---coinciding with the value for the Maxwell theory shown in green--- and its upper bound is $C_{\ssc T}/F_0\big|_{U(N)_k\times U(N)_{-k}\text{ ABJM}}\leq3/(2\pi^2\log 4k)$. As $N$ increases, the upper bound tends to the holographic value \eqref{holoE} ---shown in gray. For clarity we also show the quotient corresponding to the free scalar}\label{fig:ABJM}
\end{figure}

%We can also check the bound in the  $U(N)_k\times U(N)_{-k}$ ABJM theory \cite{Aharony:2008ug}, where the Chern-Simons level takes values $k=1,2$, using results from \cite{Nishioka:2013haa,Chester:2014fya}. For $N=1$ we find
%\begin{equation}
%\frac{C_{\ssc T}^{U(1)_1\times U(1)_{-1}\text{ ABJM}}}{F_0^{U(1)_1\times U(1)_{-1}\text{ ABJM}}}\simeq0.109632 \, , \quad \frac{C_{\ssc T}^{U(1)_2\times U(1)_{-2}\text{ ABJM}}}{F_0^{U(1)_2\times U(1)_{-1}\text{ ABJM}}}\simeq 0.073088\, ,
%\end{equation}
We can also check the bound in the  $U(N)_k\times U(N)_{-k}$ ABJM theory \cite{Aharony:2008ug}. %, where the Chern-Simons level takes the values $k=1,2,\ldots$
 Using results from \cite{Kapustin:2010xq,Marino:2011eh,Nishioka:2013haa,Chester:2014fya}, for $N=1$ we find
\begin{equation}
\cTFb{U(1)_k\times U(1)_{-k}\text{ ABJM}}\leq\frac{3}{2\pi^2\log 4k}\simeq0.109632 \, ,
\end{equation}
where $\cTb{U(1)_k\times U(1)_{-k}\text{ ABJM}}$ does not depend on $k$. Observe that at large $k$, the quotient tends to 0, saturating the lower bound \eqref{ff04}. For non-Abelian ABJM theories, the explicit expressions of $C_{\ssc T}$ and $F_0$ become more complicated. Using those computed in \cite{Chester:2014fya} for $N=2$ we find
\begin{equation}
\cTFb{U(2)_k\times U(2)_{-k}\text{ ABJM}}\lesssim 0.0905273\, .
\end{equation}
where, again, as $k$ grows, the quotient approaches 0. In both instances, we observe that the bound is satisfied. We also checked explicitly that this is true for $N=3,4$ with $k=3$, whose expressions for $C_{\ssc T}$ and $F_0$ can be found in \cite{Chester:2014fya,Binder:2020ckj,Alday:2021ymb}. On the other hand, in the large $N$ limit, the values of both $C_{\ssc T}$ and $F_0$ are expressed in terms of the Airy function \cite{Codesido:2014oua,Agmon:2017xes,Bobev:2022eus}. In this case, the quotient tends to the holographic one \eqref{holoE}, as checked for several values of small and large $k$. In Figure \ref{fig:ABJM} we plot some values of $C_{\ssc T}/F_0\big|_{U(N)_k\times U(N)_{-k}\text{ ABJM}}$ corresponding to the cases mentioned above. 

\subsection{Holographic higher-curvature theories}
Consider a holographic higher-curvature gravity in four dimensions defined by a Lagrangian density 
\begin{equation}
\mathcal{L}(R_{\mu\nu\rho\sigma},g_{\mu \nu})= \frac{1}{16\pi G} \left[ \frac{6}{L^2}+R+\sum_{n=2} \mu_n L^{2(n-1)} \mathfrak{R}_{(n)}\right]\, ,
\end{equation}
where $L$ is a length scale, $G$ is the Newton constant, $\mu_n$ are dimensionless  couplings and $\mathfrak{R}_{(n)}$ are higher-curvature densities of order $n$. %Then, the vacuum solutions can be obtained by solving
%\begin{equation}
%\Upsilon
%\end{equation}
The result for $F_0$ can be obtained by evaluating the Lagrangian on the AdS background as \cite{Imbimbo:1999bj,Schwimmer:2008yh,Myers:2010tj,Bueno:2018xqc}
%$\mathcal{L}(R_{\mu\nu\rho\sigma},g_{\mu \nu})$, $F_0$ is given by
\begin{equation}
F_0=-\frac{4\pi^2 L_{\star}^4}{3} \mathcal{L}(x)\, ,
\end{equation}
where $\mathcal{L}(x)$  stands for the on-shell Lagrangian of the corresponding theory evaluated on pure AdS with radius $L_{\star}$, which we expressed in terms of the quotient  $x\equiv L^2/L_{\star}^2$. The AdS radius and the action scale $L$ are related on-shell by the equation
%where the AdS radius $L_{\star}$ is related to the action scale $L$ through the equation 
\cite{Aspects}
\begin{equation}\label{osADS}
 \mathcal{L}(x) = \frac{x}{2} \mathcal{L}'(x) \, ,
\end{equation}
where, in order to take the derivative with respect to $x$, $ \mathcal{L}(x)$ must be expressed in terms of powers of $x $ up to an overall $1/L^2$.
%where the on-shell Lagrangian has been written in terms of the dimensionless ratio $x\equiv L^2/L_{\star}^2$ before taking the derivative in the second term.
For instance, for pure Einstein gravity one has $\mathcal{L}(x)=3(1-2x)/(8\pi L^2 G)$ and $x\mathcal{L}'(x)/2=-3x/(8\pi L^2 G)$ so \req{osADS} simply implies $x=1$, \ie $L=L_{\star}$.

On the other hand, for higher-curvature theories which have second-order equations on maximally symmetric backgrounds, $\ctt$ can be computed as \cite{Bueno:2018yzo}
\begin{equation}
\ctt =- \frac{4L^2 L_{\star}^2}{\pi^2} \left[ \mathcal{L}'(x)-x \mathcal{L}''(x) \right]\, .
\end{equation}
%For instance, for Einstein gravity one has $ \mathcal{L}''(x)=0$, $\ctt =- 4L^2L_{\star}^2/\pi^2 $
%is the Einstein gravity result.
 Hence, for the ratio of interest we find
\begin{equation}
\frac{\ctt}{F_0}=\frac{6}{\pi^4} \left[1-\frac{ x \mathcal{L}''(x)}{\mathcal{L}'(x)} \right] \, , 
\end{equation}
where we made use of \req{osADS}. Naturally, for pure Einstein gravity, this reduces to \req{holoE}, since in that case $ \mathcal{L}''(x)=0$. Using our conjectural bounds (\ref{ff04}), one finds the following conditions on the higher-curvature couplings of this class of theories
\begin{equation}
 -1.4168 \lesssim \frac{ x \mathcal{L}''(x)}{\mathcal{L}'(x)} \leq 1\, .
\end{equation}
More generally, the bounds can be used to put constraints on the possible values of the higher-curvature couplings of other holographic theories for which $\ctt$ and $F_0$ are available ---\eg \cite{Bobev:2021oku,Bobev:2020egg}. Note however that the meaning of such bounds is somewhat unclear since, generically, higher-curvature theories give rise to instabilities  when considered beyond the perturbative effective field theory regime.

\bibliography{Gravities}
\bibliographystyle{apsrev4-1} % Tell bibtex which bibliography style to use
\vspace{1cm}

\end{document}
%
% ****** End of file template.aps ******